\documentclass[draft]{agujournal2019}
\usepackage{url} 
\usepackage{lineno}
\usepackage[inline]{trackchanges} 
\usepackage{soul}
\draftfalse

\journalname{Geophysical Research Letters}

\begin{document}

%
%


\title{Applying Magnetic Curvature to to MMS data to identify thin current sheets relative to tail reconnection}

%
%




\authors{A.J. Rogers\affil{1}, C.J. Farrugia\affil{1}, R.B. Torbert\affil{1}, T.J. Rogers\affil{2}}


\affiliation{1}{Space Science Center, University of New Hampshire, Durham, New Hampshire}
\affiliation{2}{Microsoft Corporation, Redmond, Washington}




\correspondingauthor{Anthony Jason Rogers}{anthony.rogers@unh.edu}




\begin{keypoints}
\item Geomagnetic tail current sheet thickness during MMS 2017--2020 tail seasons estimated using magnetic field line curvature
\item Location of tail current sheets with thickness at or below ion gyro-scale compared to location of reconnection-related Ion Diffusion Regions
\item Observations compared to recent PIC simulations
\end{keypoints}

%
%

%
%


\begin{abstract}
Magnetic reconnection X-lines have been observed to be more common duskward of midnight. Thin current sheets have also been postulated to be a necessary precondition for reconnection onset.  We take advantage of the MMS tetrahedral formation during the 2017--2020 MMS tail seasons to calculate the thickness of the cross-tail neutral sheet relative to ion gyroradius. While a similar technique was applied to Cluster data, current sheet thickness over a broader range of radial distances has not been robustly explored before this study. We compare this to recent theories regarding mechanisms of tail current sheet thinning and to recent  simulations. We find MMS spent more than twice as long in ion-scale thin current sheets in the pre-midnight sector than post-midnight, despite nearly even plasma sheet dwell time. The dawn-dusk asymmetry in the distribution of Ion Diffusion Regions, as previously reported in relation to regions of thin current sheets, is also analyzed.
	

\end{abstract}

\section*{Plain Language Summary}
Magnetic reconnection is an important mechanism for energy transfer in the magnetosphere.  In order for reconnection to begin, however, the reconnecting current sheet must first become very thin.  In the geomagnetic tail reconnection and related phenomena have been observed closer to dusk than dawn on the nightside, although the reasons for this have not been clearly understood.  Recent simulations of the geomagnetic tail suggest that the central current sheet in the tail should be thinner pre-midnight than post-midnight, possibly explaining why reconnection happens more often on the pre-midnight than the post-midnight sector.  We use nineteen months of MMS data in the tail, comprising the tail seasons of four years from 2017--2020, to estimate the thickness of the tail neutral sheet relative to relevant ion scales from dawn flank to dusk flank and both closer to and further away from the Earth than has been done in the past.  We then compare the thickness we measure with the simulation predictions and with the location of previously identified reconnection locations in the same time period.

%
%

%


%
%
%
%

\section{Introduction}

	In Dungey's model of the open magnetosphere (1961, 1963), energy stored in the interplanetary magnetic field (IMF) is transferred to kinetic energy of the magnetospheric plasma in a process called magnetic reconnection. On the dayside, reconnection is between the IMF and that of the closed terrestrial magnetospheric field lines. In the tail, where the flux transported from the dayside loads and thins the tail current sheet (CS, embedded in the plasma sheet), reconnection closes the open flux and returns it to the dayside, leading to a two-cell plasma convection (see review by Cowley, 1981).  This model has received abundant observational support.

	Phenomena associated with reconnection in the geomagnetic tail such as auroral substorms, dipolarization fronts, and bursty-bulk flows (BBFs) as well as \textit{in situ} observations of reconnection in the tail have provided an important part of this support.  However, the locations of these phenomena have also shown a significant asymmetry in the dawn-dusk direction, being more common on the dusk-side of midnight by significant ratios (Nagai et al., 1998; Posch et al., 2007; Xiao et al., 2017).  A statistical study using THEMIS (Sibeck and Angelopoulous, 1998) was performed by Imber et al. (2011) and showed that 81\% of flux ropes and traveling compression regions associated with reconnection in the magnetotail were found in the dusk sector.  Rogers, Farrugia, and Torbert (2019) showed that reconnection regions (ion diffusion regions or IDRs) during the 2017 MMS tail season (May -- September) were preferentially observed \textit{in situ} on the dusk-side of midnight by a far larger ratio (91.7\% pre-midnight, 8.3\% post-midnight) than could be accounted for by differences in spacecraft dwell time on either side of midnight (56.5\% pre-midnight, 43.5\% post-midnight).  This suggests that the observed asymmetrical distribution of reconnection and related phenomena is due to underlying physical processes and not observational bias.
	
	Reconnection theory (S\"onnerup, 1979) and simulations (e.g. Birn, 1980; Liu et al., 2019) suggest that a thin current sheet, \textit{i.e.} one with a thickness on the order of an ion inertial length ($d_{i}$) or less, is necessary for reconnection onset to occur.  Ion-scale thin current sheets in the neighborhood of tail reconnection sites have been anecdotally observed (e.g. Runov et al., 2003), which supports a correlation between extended regions of thin current sheets and magnetic reconnection in the geomagnetic tail.  The distribution of ion-scale current sheets within the central tail plasma sheet is thus pertinent to the question of reconnection location in the tail.
	
	Attempts to estimate the thickness of the tail current sheet have been occasionally made using single-spacecraft measurement techniques (\textit{e.g.} Artemyev et al., 2011; S. Lu et al., 2019).  Many of these techniques relied upon a ratio of measurements of the magnetic field and particle current density.  These were sometimes made at substantially different points in time, which assumes that the current sheet being measured is essentially a quiet, Harris-type current sheet (Harris, 1962) throughout the measurement period.  However, reconnection and related substorm phenomena often occur during times of a disturbed geomagnetic field, violating the assumption of a Harris-like current sheet.
	
	Rong et al. (2011) utilized the four spacecraft of the Cluster mission in tetrahedral formation to calculate the radius of curvature ($R_{C}$) of the magnetic field at the barycentre of the fleet:
	\begin{equation}
		R_{C} = \frac{1}{|(\hat{b} \cdot \nabla)\hat{b}|}
	\end{equation}
	where $\hat{b}$ is the unit vector $\vec{B}/|\vec{B}|$.  They then estimated the thickness of the neutral sheet by scaling the radius of curvature with the current sheet tilt, as described by Shen et al. (2008).  This method was superior to previous estimates using single-spacecraft techniques. A statistically significant dawn-dusk asymmetry in current sheets thinner than 1000km was found, with thinner currents sheets more common duskward of midnight.  However, due to the nature of its near-polar orbit, Cluster only encountered the tail current sheet at a radial distance approximately 20 Earth radii ($R_{E}$) from the Earth within a narrow band $\approx 2 R_{E}$ wide.  Current sheet thickness in the geomagnetic tail has not been explored using such robust techniques at other radial distances before our current study.
	
	Current sheet thinning has also been addressed in numerical simulations.  Recent hybrid (Lu et al., 2016) and particle-in-cell (PIC) simulations (Lu et al., 2018) suggest a mechanism for preferential thinning of the tail current sheet in the pre-midnight sector.  In both simulations, external drivers cause a global thinning of the tail current sheet to approximately ion scales, at which point partial demagnetization of ions drives charge separation from the still frozen-in electrons, leading to Hall electric fields.  The duskward cross-tail current of the tail current sheet is then enhanced by the $\vec{E} \times \vec{B}$ drifting of frozen-in electrons and diamagnetic drift of partially demagnetized ions, leading to progressive thinning of the current sheet on the pre-midnight.  These simulations suggest the asymmetric thinning should be robust across a broad range of radial distance. 
	
	The MMS mission (Burch et al. 2015) launched in 2015 typically flies in a tetrahedron formation that is capable of the 4-point measurements necessary for curvature calculation. We utilize a technique particularly suited to identifying current sheets which may be preferentially thinned as suggested by simulations.  Bu\"chner and Zelenyi (1989) described a scalar parameter $K$ to identify whether the local magnetic field could support adiabatic motion of charged particles in a plasma.  We adapt $K$ to apply this test against the thermal ion population in the vicinity of the MMS observatories:
	\begin{equation}
		K_{i} = \sqrt{\frac{R_{C}}{\rho_{g,i}}}
	\end{equation}
	where $\rho_{g,i} = \frac{\sqrt{2m_{i}T_{\perp,i}k_{b}}}{|q_{i}||\vec{B}|}$ is the thermal ion gyroradius.  Where $K_{i} < 1$ the gyroradius of the ions is larger than the radius of curvature of the local magnetic field, implying that the majority ions will not remain frozen-into the magnetic field and  adiabatic motion of those ions cannot be supported.  Under such conditions Hall electric fields may form between the demagnetized ions and the still-frozen-in electrons, supporting the process described by Lu et al.
	
	We calculate the value of $K_{i}$ throughout the geomagnetic tail near the plasma sheet for the entire 2017--2020 MMS tail seasons (Phases 2b, 3b, 4b, and 5b).  As $K_{i}$ is a measure of the relative thickness of the neutral sheet, recalling that $R_{C}$ is proportional to neutral sheet thickness (see section 3.8 Rong et al. 2011), we use these data to test recent theories regarding mechanisms for causing thinning of the tail current sheet by   comparing MMS observations to predictions made in simulations.  We also investigate the relationship between thin (ion gyroradius scale) current sheet locations and the          occurrence of Ion Diffusion Regions (IDRs) associated with reconnection.

%
%
%
%
%
%
%
%
%
%
%
%
%

\section{Instruments and methods}

The MMS spacecraft measure electric and magnetic fields using the FIELDS instrument suite (Torbert et al. 2016).  The analog and digital fluxgate magnetometers (AFG/DFG)    measure magnetic fields in the frequency range from DC up  to 64 Hz (Russell et al. 2016).  Level 2 fluxgate magnetometer (FGM) data of version 5.86 and higher (highest     available as of submission) were used throughout this study.

The Fast Plasma Instrument (FPI) provides MMS with high cadence electron and ion distributions in the energy/charge range of 10 eV/q up to 30 keV/q. Each MMS satellite is   equipped with eight FPI         spectrometers which, combined with electrostatic control of the field-of-view, allows FPI to sample the full electron and ion distributions  (Pollock et al. 2016). It is important to note that core ion distributions can extend beyond the range of FPI, meaning that actual ion temperatures may be higher than what  is calculated using FPI data.  Level 2 FPI ion moments of version 3.3.0 were used throughout this study.

Positions of the individual spacecraft in the MMS fleet are provided using Magnetic Ephemeris and Coordinates (MEC) data products (Morley, 2015) and are calculated using    the LANLGeoMag suite (Henderson et al. 2018).  All instrument data used in this study is available from the MMS Science Data Center (https://lasp.colorado.edu/mms/sdc).     Level 2 fast survey data was used throughout this study.  Calculations of the magnetic field line curvature were made using the mms-curvature library and is publicly        available (https://github.com/unh-mms-rogers/mms-curvature).

In order to ensure that the formation of the MMS fleet was appropriate for the calculation of spatial gradients a minimum value of the Tetrahedron Quality Factor (TQF: Fusilier  et al. 2016) was required where $TQF \geq 0.8$.  For similar reasons, this survey was performed only on data collected while the MMS fleet was at least 8 Earth radii        ($R_{E}$) from the Earth to avoid deformations of the regular tetrahedron as the fleet approached perigee.  Observations in this study were also limited to regions with a measured ion density of $\rho_{i} \geq 0.05 cc^{-1}$ to ensure observations were made in the plasma sheet (Rogers, Farrugia, Torbert, 2019; Raj et al. 2002; Baumjohann, 1993).  

\begin{figure}[!h]
	\centering
	\includegraphics[width=\textwidth]{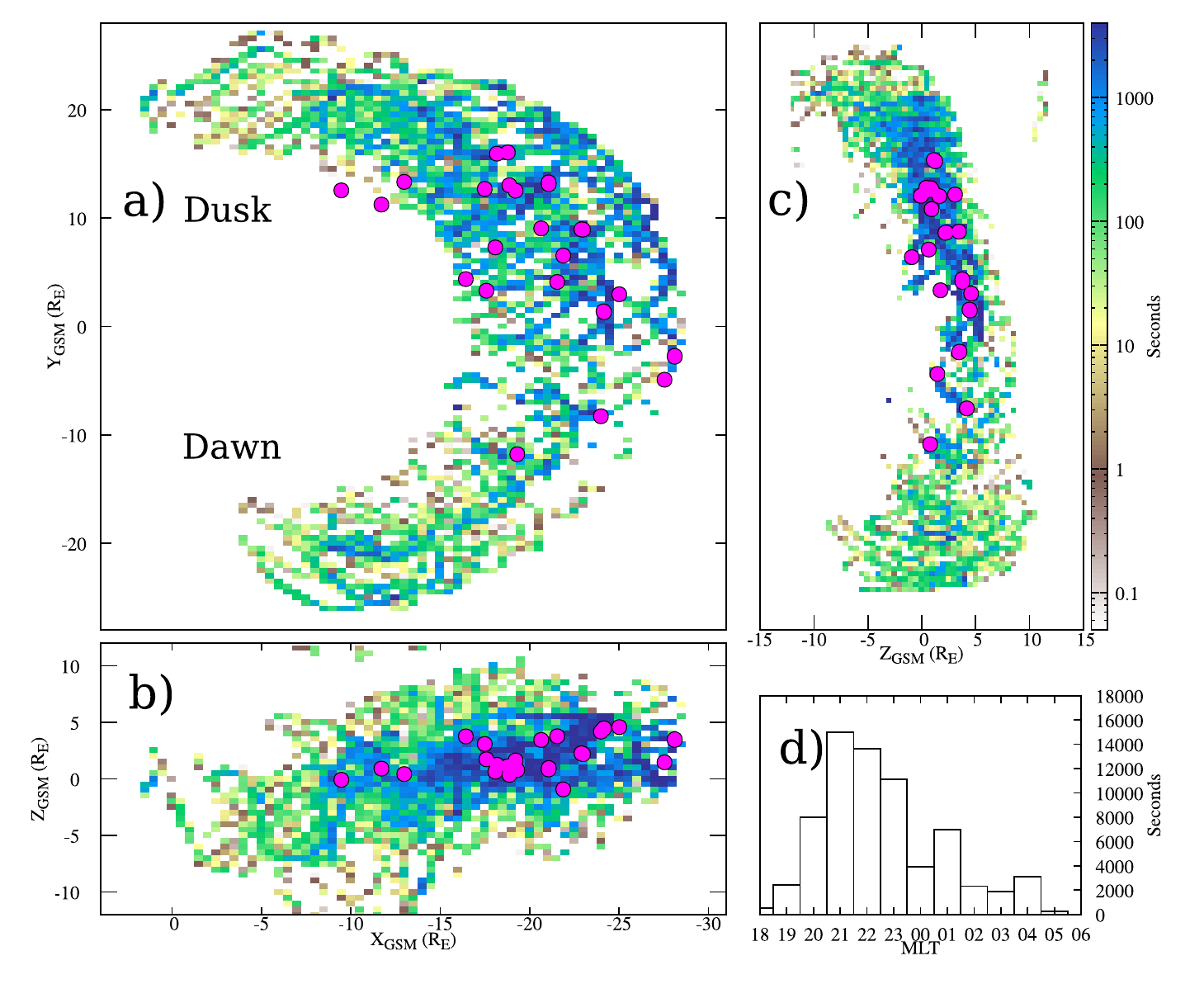}
	\caption{The total amount of time in seconds spent by MMS in a TCS, under conditions outlined in the text, in sectors of dimension $0.5 R_{E} \times 0.5 R_{E}$ (\textit{a, b, c}) and Magnetic Local Time (MLT) (\textit{d}).  \textit{a)} Distribution of TCS dwell times in the GSM \textit{X--Y} plane, with times summed over the GSM \textit{Z} axis.  \textit{b)} TCS dwell times in the GSM \textit{X --Z} plane, with times summed over the GSM \textit{Y} axis.  \textit{c)} TCS dwell times in the GSM \textit{Z -- Y} plane, with times summed over the GSM \textit{X} axis.  \textit{d)} TCS dwell times across all radial distances by MLT location.  Magenta circles represent Ion Diffusion Region locations (see text)}
	\label{fig:TCS-full}
\end{figure}

\section{Observations and Analysis}

Figure 1 shows the distribution of ion-scale thin current sheets as measured by MMS over the combined 2017--2020 tail seasons where data from all four spacecraft were available ($\approx$ 550 days).  Colors represent the amount of time which MMS spent in each region with a value of $K < 1.0$, \textit{i.e.} the dwell time of MMS in a thin current sheet (TCS).  The effects of orbital bias on these dwell time measurements are small as MMS spent approximately equal time in the plasma sheet on either side of midnight: 51.7\% pre-midnight, 48.3\% post-midnight. IDRs previously identified by Rogers, Farrugia, and Torbert (2019) from the 2017 season (Phase 2b), as well as additional IDRs identified using the same technique, have been overlaid as magenta circles on the TCS dwell times in Figure 1.  All of these IDRs are found in regions where MMS spent at least a moderate amount of time in a TCS.  It should be noted that some IDR markers in Figure 1 and following figures totally obscure the dwell time indicator for the region where they are located.  A listing of all identified IDRs shown here is provided in the supplementary materials. 

Figure 1d shows the total TCS dwell time as a function of Magnetic Local Time (MLT).  This sub-figure is comparable with similar plots showing the global MLT distribution of other substorm-related phenomena such at Pi1B pulsations (Posch et al., 2007, Fig.12) and dipolarization fronts (Xiao et al., 2017, Fig. 6), all of which show a strong preference for substorm-related activity to occur duskward of midnight.   The sharp delineation at midnight shown in Figure 1d is similar to distributions of substorm related phenomena reported in the previous studies mentioned.

Figure 2 shows the same TCS data derived from the parameter $K$ but here normalized by the total dwell time MMS spent in the tail plasma sheet, as indicated by the measured ion density     $n_{i} > 0.05 cc^{-1}$.  As expected, the majority of the time spent by MMS in the plasma sheet was not near an ion-scale TCS, as indicated by the bulk of the distribution  reflecting a ratio of TCS time to plasma sheet dwell time of much less than one.  The distribution of known IDRs during the same time frame is also laid over the map of     normalized TCS dwell times in Figure 2.

\begin{figure}[!h]
	\centering
	\includegraphics[width=0.8\textwidth]{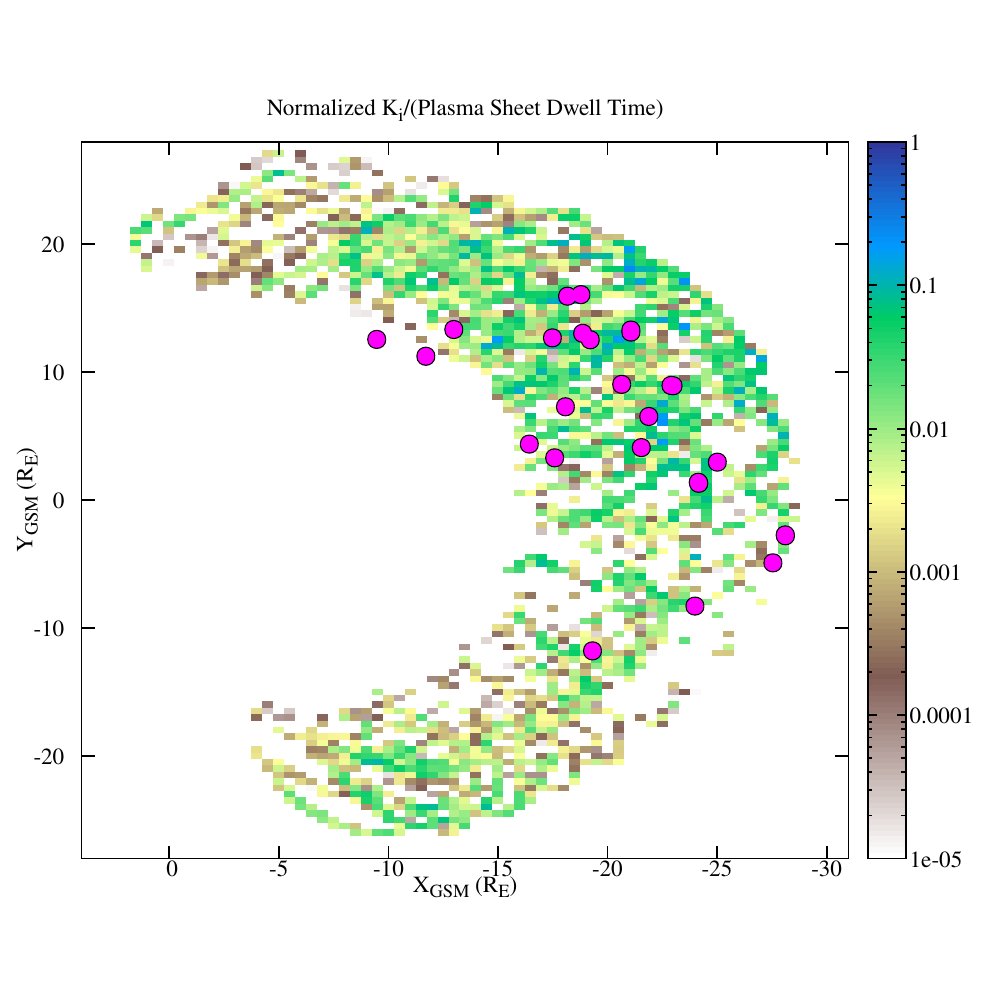}
	\caption{Dwell time MMS spent in a TCS normalized by the amount of time MMS spent in the plasma sheet.  Regions and projection are in the style of Figure 1a.}
	\label{fig:TCS-N2PS}
\end{figure}

The total time spent in a TCS is significantly higher on the pre-midnight of the tail ($187.8 hrs$) than the post-midnight ($77.68 hrs$) (see Figure 1d).  While MMS spent an approximately equal amount of time in the plasma sheet on both the pre- and post-midnight sectors ($51.7\% / 48.3\%$), this contrasts with the uneven amount of time spent in a thin current sheet by MMS; $70.7\%$ of total time spent in    TCSs was in the pre-midnight sector versus $29.3\%$ post-midnight.  The majority of the time spent in a TCS on the post-midnight is found near apogee across all seasons (see Fig.1) where spacecraft velocity    was lowest and the bulk of dwell time each orbit was spent regardless of other factors.

The distribution of time spent in a TCS pre-midnight is far more varied in radial distance and is not confined to the apogee of each orbit.  Figure 3 shows the dwell time of MMS in a TCS relative to dwell time in moderate-to-high ion density as a function of MLT in bands of radial distance, each $2 R_{E}$ wide.  The center of the relative TCS dwell time distribution is at $\approx$21 MLT at smaller radial distances (Fig. 3a,b).  The distribution both broadens towards midnight as MMS increases in radial distance (Fig. 3c,d), although the peak remains more duskward.  The large relative TCS dwell times at the dawn and dusk flanks (Fig.2) are interpreted as encounters with the flank magnetopause where current sheets and increased ion density are expected.  Sub-figure $3c)$ represents the approximate region of the geomagnetic tail sampled by Cluster as in Rong et al. (2011).

The asymmetry in the locations of Ion Diffusion Regions (IDRs; magenta circles, Figure 1) associated with magnetic reconnection is even more pronounced than that of relative TCS dwell time.  19 IDRs were confidently identified across all four seasons on the dusk-side of midnight, while only six were observed on the dawn-side, equating to 76.9\% of reconnection events observed duskward of midnight and 23.1\% dawnward.  In Figure 2 we see that the majority of previously identified IDRs lie not only in areas where MMS spent a great deal of time in ion-scale TCS, but also where the ratio of TCS dwell time to total plasma sheet dwell time was greatest.  This supports the intuitive interpretation that a thinner current sheet is more likely to support reconnection, and that reconnection is more likely where thin current sheets are more common.  However, the long periods of time spent in both the plasma sheet and thin current sheet where no reconnection was observed, such as -16$R_{E} \hat{x}$, 11$R_{E} \hat{y}$ in Figure 2, indicate that an ion-scale TCS is an insufficient condition for reconnection.


\begin{figure}[!h]
	\centering
	\includegraphics[width=0.9\textwidth]{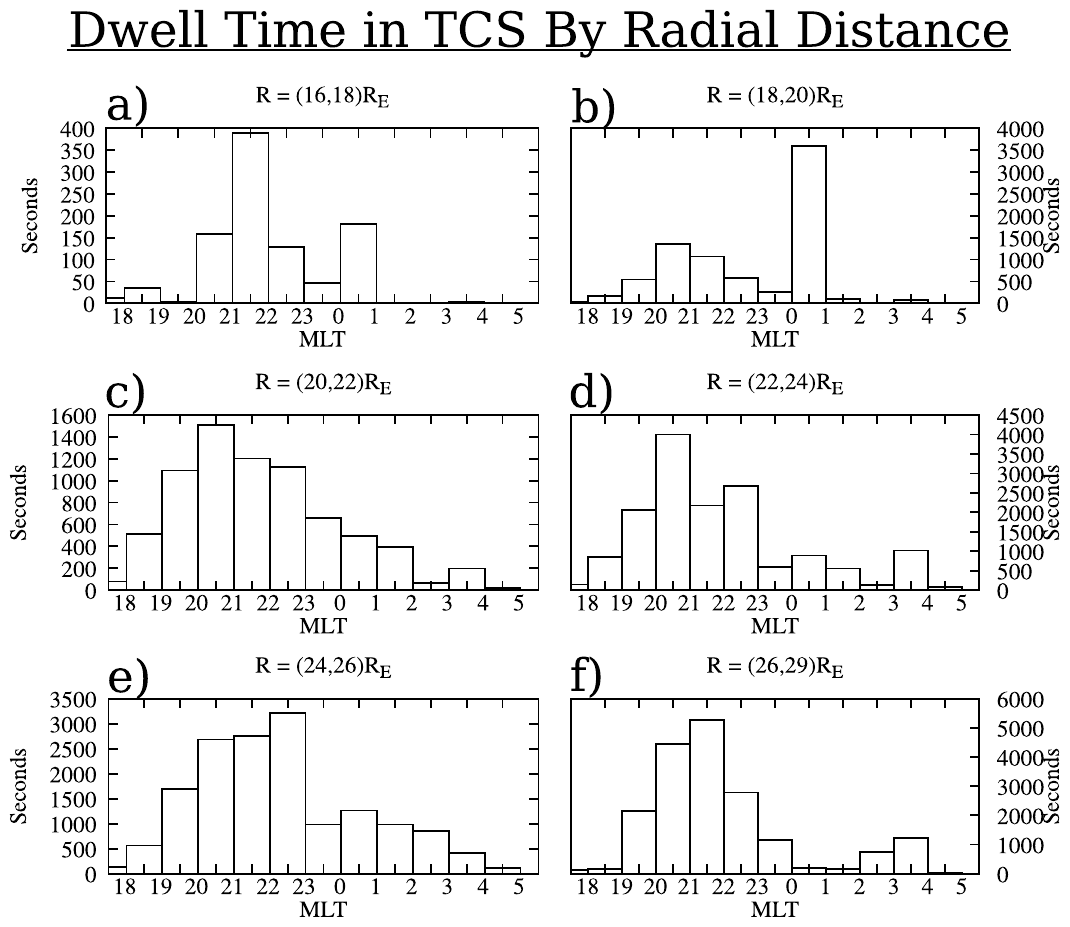}
	\caption{Dwell time MMS spent in a TCS by MLT for narrow ($2 R_{E}$ wide) bands across the geomagnetic tail, normalized by the total amount of time MMS spent in the plasma sheet in that same MLT $\times$ R sector.  \textit{e)} Is the approximate band which has been studied by the Cluster mission (\textit{e.g.} Rong et al. 2011).}
	\label{fig:TCS_by_R}
\end{figure}

\section{Discussion}

A feature of tail reconnection that stand in observations is that IDRs and related reconnection phenomena occurs preferentially in the pre-midnight sector (Eastwood et al. 2010,      Nagai et al. 2013, Genestreti et al. 2013).  The explanation for this has been glossed over somewhat since the question of observational bias due to orbital variations was  not often addressed (see discussion in Imber et al. 2011).  If a spacecraft spends more time at dusk then one supposes its chances of seeing IDRs or other phenomena is increased relative to other    regions, all other things being equal.  Thus while studies based on previous missions showed a dawn-dusk asymmetry in observed reconnection-related features, we could not be  sure it was not due to an observational bias.  This question was addressed by Rogers, Farrugia, and Torbert (2019) who confirmed that the asymmetric distribution of observed IDRs by MMS    was not a function of observational bias but is a result of magnetotail physics.  

The strong asymmetry in typical current sheet thickness is not entirely surprising.  In addition to previous studies which observed a similar asymmetry but were more        limited in extent (Rong et al. 2011) or utilized more indirect methods of calculating current sheet thickness (e.g. S. Lu et al. 2019), mechanisms for the source of this    asymmetry have also been mooted.   Lu et al. (2018) and Pritchett and Lu (2018) have hypothesized that the cause of this dusk-preference was preferential thinning on the pre-midnight of the tail due an externally-driven convective electric field and enhanced by Hall electric fields (normal to the current sheet)     which formed as the current sheet approached ion scales.  Comparing our observations to this model we find the ratio of dwell time in an ion-scale TCS, 70.7\% pre-midnight to 29.3\% post-midnight; a 2 to 1 ratio with twice as much time spent encountering an ion-scale current sheet pre-midnight as post-midnight, qualitatively supports the model of Lu et al.  A comparison of our Figures \ref{fig:TCS-N2PS} and \ref{fig:TCS_by_R} to Figure 2 in Lu et al. (2018) is even more encouraging as the radial variations in our observations are qualitatively similar to those shown in the PIC simulations.  



\begin{figure}[!h]
	\centering
	\includegraphics[width=0.4\textwidth]{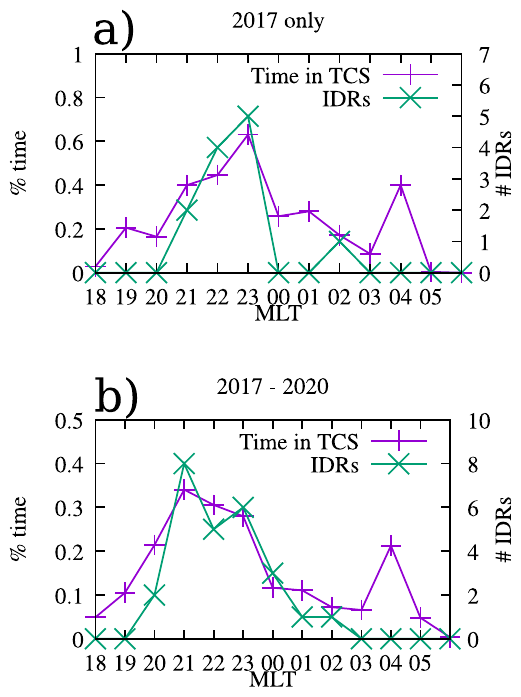}
	\caption{Plots of the proportion of the time MMS spent in a TCS relative to total time spent in the plasma sheet as well as the location of identified IDRs, both as a function of MLT, for the 2017 tail season (\textit{a}) and the combined 2017--2020 tail seasons (\textit{b}).}
	\label{fig:time_idr_mlt}
\end{figure}
	
We can also see the pre-/post-midnight TCS asymmetry clearly in Figure \ref{fig:time_idr_mlt} which plots the time MMS spent in a TCS as a percentage of the total time spent in the plasma sheet, along with the number of IDRs identified in each MLT region.  It is unsurprising that the number of IDRs identified increases with the greater proportional time spent in a TCS, but less expected is the greater duskward extent of the TCS than observed IDRs.  

Recent 3D PIC simulations (Liu et al. 2019) suggest that the near-midnight preference of observed reconnection may be due to the effects of Hall reconnection in three dimensions.  Liu et al. suggest that reconnection on a given thin current sheet will be suppressed on the side of the current sheet opposite the direction of net current flow for a region on the scale of 10s of ion inertial lengths (see Fig. 7, Liu et al., 2019).  In the geomagnetic tail, with a net dawnward flow, this would lead to suppression of reconnection onset on the duskward side of an existing thin current sheet.  As we have observed, ion-scale TCSs capable of supporting possible reconnection are twice as likely to be found on the pre-midnight of midnight as on the post-midnight.  Thus while reconnection is more likely to occur duskward of midnight in the tail due to ion-scale TCSs occurring more frequently there, it is more likely to be suppressed as the TCS extends further duskward away from midnight (see Fig. 9, Liu et al., 2019).  This would have the practical effect of concentrating reconnection onset in the region at and near midnight in the pre-midnight region.  The location of nearly half of all identified IDRs within the region of 22-24MLT supports this behavior.

Also of note is the duskward expansion of both TCS and IDR locations over the four tail seasons of this study.  2017-2019 were in the declining phase of the solar cycle.   Figure \ref{fig:time_idr_mlt}\textit{a} shows the distribution of proportional time in a TCS as well as IDR distribution in MLT for the 2017 tail season with a 2017 annual average Dst of $-8.09 nT$.  Figure \ref{fig:time_idr_mlt}\textit{b} shows the same distributions for the full four seasons of this study; during which time the Dst decreased to an annual average of $-5.14 nT$ in 2019 and averaged only $-5.54 nT$ for the whole of 2018-2020 (Nose et al. 2015).  This implies not only that the process of plasma sheet compression is a global process dependant on solar activity, known since Dungey (1963), but that the mesoscale process of asymmetric TCS distribution is also a function of solar driving.  More observations of tail TCS distributions during the increasing solar cycle may provide more insight into this question, or at least provide better statistics.


\section{Conclusion}

Locations in the geomagnetic tail where the neutral sheet thickness is reduced to ion scales have been mapped during four MMS tail seasons (2017--2020) using the ratio of radius of magnetic curvature to the thermal ion gyroradius.  The routine calculation of magnetic field curvature was made possible due to the high-resolution magnetic field          measurements available on all four spacecraft of the MMS fleet, as well as the regular tetrahedron geometry of their formation.  Ion-scale thin current sheets were found to be twice as common on the pre-midnight side of the geomagnetic tail as on the dawn-side.  Locations of common thin current sheets were compared to the distribution of reconnection  Ion Diffusion Regions previously identified for the same time span and implications for their coincidence were discussed.  Possible mechanisms for the formation of both     thin current sheets and reconnection suggested by recent PIC simulations were compared to these observations and qualitative agreement with simulations was found.

\acknowledgments
The authors would like to thank Zach Dykstra and Phil Doroff For computational support.  All instrument data used in this study is publically available at the MMS Science Data Center (https://lasp.colorado.edu/mms/sdc/public/).  This work has been supported by NASA via contract 499878Q


%
%

\section*{References}
\subsection*{--- To be replaced with BibTeX later ---}

Artemyev, A. V., et al. “Cluster Statistics of Thin Current Sheets in the Earth Magnetotail: Specifics of the Dawn Flank, Proton Temperature Profiles and Electrostatic Effects.” J. Geophys. Res. Sp. Phys., vol. 116, no. 9, 2011, pp. 1–9, doi:10.1029/2011JA016801.

Baumjohann, Wolfgang. “The Near-Earth Plasma Sheet: An AMPTE/IRM Perspective.” Space Science Reviews, vol. 64, no. 1–2, 1993, pp. 141–63. DOI.org (Crossref), doi:10.1007/BF00819660.

Birn, J. “Computer Studies of the Dynamic Evolution of the Geomagnetic Tail.” Journal of Geophysical Research: Space Physics, vol. 85, no. A3, Mar. 1980, pp. 1214–22. DOI.org (Crossref), doi:10.1029/JA085iA03p01214.

Büchner, Jörg, and Lev M. Zelenyi. “Regular and Chaotic Charged Particle Motion in Magnetotaillike Field Reversals: 1. Basic Theory of Trapped Motion.” J. Geophys. Res., vol. 94, no. A9, 1989, pp. 11821–42, doi:10.1029/ja094ia09p11821.

Burch, J. L., et al. “Magnetospheric Multiscale Overview and Science Objectives.” Space Science Reviews, vol. 199, no. 1–4, Mar. 2016, pp. 5–21. DOI.org (Crossref), doi:10.1007/s11214-015-0164-9.

Dungey, J.W. “Interactions of Solar Plasma with the Geomagnetic Field.” Planetary and Space Science, vol. 10, Jan. 1963, pp. 233–37. DOI.org (Crossref), doi:10.1016/0032-0633(63)90020-5.

Dungey, J. W. “Interplanetary Magnetic Field and the Auroral Zones.” Phys. Rev. Lett., vol. 6, no. 2, 1961, pp. 47–48, doi:10.1103/PhysRevLett.6.47.

Eastwood, J. P., et al. “Average Properties of the Magnetic Reconnection Ion Diffusion Region in the Earth’s Magnetotail: The 2001-2005 Cluster Observations and Comparison with Simulations.” J. Geophys. Res. Sp. Phys., vol. 115, no. A08215, 2010, doi:10.1029/2009JA014962.

Fuselier, S. A., et al. “Magnetospheric Multiscale Science Mission Profile and Operations.” Space Science Reviews, vol. 199, no. 1–4, Mar. 2016, pp. 77–103. DOI.org (Crossref), doi:10.1007/s11214-014-0087-x.

Genestreti, K. J., et al. “The Location and Rate of Occurrence of Near-Earth Magnetotail Reconnection as Observed by Cluster and Geotail.” J. Atmos. Solar-Terrestrial Phys., vol. 121, 2014, pp. 98–109, doi:10.1016/j.jastp.2014.10.005.

Harris, E. G. “On a Plasma Sheath Separating Regions of Oppositely Directed Magnetic Field.” Il Nuovo Cimento, vol. 23, no. 1, Jan. 1962, pp. 115–21. DOI.org (Crossref), doi:10.1007/BF02733547.

Henderson, Mike, et al. LANLGeoMag: V1.5.16 (Version v1.5.16). Zenodo, 10 Mar. 2018, http://doi.org/10.5281/zenodo.1195041.

Imber, S. M., et al. “A THEMIS Survey of Flux Ropes and Traveling Compression Regions: Location of the near-Earth Reconnection Site during Solar Minimum.” J. Geophys. Res. Sp. Phys., vol. 116, no. A02201, 2011, p. 12, doi:10.1029/2010JA016026.

Liu, Yi Hsin, et al. “Three-Dimensional Magnetic Reconnection With a Spatially Confined X-Line Extent: Implications for Dipolarizing Flux Bundles and the Dawn-Dusk Asymmetry.” J. Geophys. Res. Sp. Phys., vol. 124, no. 4, 2019, pp. 2819–30, doi:10.1029/2019JA026539.

Lu, San, P. L. Pritchett, et al. “Formation of Dawn-Dusk Asymmetry in Earth’s Magnetotail Thin Current Sheet: A Three-Dimensional Particle-In-Cell Simulation.” J. Geophys. Res. Sp. Phys., vol. 123, no. 4, Blackwell Publishing Ltd, Apr. 2018, pp. 2801–14, doi:10.1002/2017JA025095.

Lu, San, Y. Lin, et al. “Hall Effect Control of Magnetotail Dawn-Dusk Asymmetry: A Three-Dimensional Global Hybrid Simulation.” J. Geophys. Res. Sp. Phys., vol. 121, no. 12, 2016, p. 11,882-11,895, doi:10.1002/2016JA023325.

Morley, Steve. Magnetic Ephemeris and Coordinates. Zenodo. doi:10.5281/zenodo.2594027

Nagai, T, et al. “Structure and Dynamics of Magnetic Reconnection for Substorm Onsets with Geotail Observations.” J. Geophys. Res., vol. 103, A3, 1998, doi:10.1029/97JA02190.

Nagai, T., et al. “Three-Dimensional Structure of Magnetic Reconnection in the Magnetotail from Geotail Observations.” J. Geophys. Res. Sp. Phys., vol. 118, 2013, pp. 1667–78, doi:10.1002/jgra.50247.

M. Nose, T. Iyemori, M. Sugiura, T. Kamei, World Data Center for Geomagnetism, Kyoto. (2015), Geomagnetic Dst index, doi:10.17593/14515-74000. 

Pollock, C., et al. “Fast Plasma Investigation for Magnetospheric Multiscale.” Space Science Reviews, vol. 199, no. 1–4, Mar. 2016, pp. 331–406. DOI.org (Crossref), doi:10.1007/s11214-016-0245-4.

Posch, J. L., et al. “Statistical Observations of Spatial Characteristics of Pi1B Pulsations.” Journal of Atmospheric and Solar-Terrestrial Physics, vol. 69, no. 15, Nov. 2007, pp. 1775–96. DOI.org (Crossref), doi:10.1016/j.jastp.2007.07.015.

Pritchett, P. L., and San Lu. “Externally Driven Onset of Localized Magnetic Reconnection and Disruption in a Magnetotail Configuration.” J. Geophys. Res. Sp. Phys., vol. 123, no. 4, 2018, pp. 2787–800, doi:10.1002/2017JA025094.

Raj, Arjun, et al. “Wind Survey of High-Speed Bulk Flows and Field-Aligned Beams in the near-Earth Plasma Sheet.” J. Geophys. Res. Sp. Phys., vol. 107, no. A12, 2002, doi:10.1029/2001JA007547.

Rogers, A. J., et al. “Numerical Algorithm for Detecting Ion Diffusion Regions in the Geomagnetic Tail With Applications to MMS Tail Season 1 May to 30 September 2017.” Journal of Geophysical Research: Space Physics, vol. 124, no. 8, Aug. 2019, pp. 6487–503. DOI.org (Crossref), doi:10.1029/2018JA026429.

Runov, A., et al. “Current Sheet Structure near Magnetic X-Line Observed by Cluster.” Geophys. Res. Lett., vol. 30, no. 11, 2003, doi:10.1029/2002GL016730.

Russell, C. T., et al. “The Magnetospheric Multiscale Magnetometers.” Space Sci. Rev., vol. 199, 2016, doi:10.1007/s11214-014-0057-3.

Sibeck, D. G., and V. Angelopoulos. “THEMIS Science Objectives and Mission Phases.” Space Science Reviews, vol. 141, no. 1–4, Dec. 2008, pp. 35–59. DOI.org (Crossref), doi:10.1007/s11214-008-9393-5.

Shen, C., et al. “Magnetic Configurations of the Tilted Current Sheets in Magnetotail.” Annales Geophysicae, vol. 26, no. 11, Nov. 2008, pp. 3525–43. DOI.org (Crossref), doi:10.5194/angeo-26-3525-2008.

Sonnerup, B. U. Ö. “Magnetic Field Reconnection.” Sol. Syst. Plasma Phys., edited by L. J. Lanzerotti et al., North-Holland Publishing Co., 1979, pp. 45–108.

Torbert, R. B., et al. “The FIELDS Instrument Suite on MMS: Scientific Objectives, Measurements, and Data Products.” Space Sci. Rev., vol. 199, 2016, doi:10.1007/s11214-014-0109-8.

Xiao, Sudong, et al. “Occurrence Rate of Dipolarization Fronts in the Plasma Sheet: Cluster Observations.” Ann. Geophys., vol. 35, no. 4, 2017, pp. 1015–22, doi:10.5194/angeo-35-1015-2017.

\bibliography{kappa_refs}

%
%
%
%
%

\end{document}